\documentclass[aps,prl,twocolumn,showpacs,psfig,superscriptaddress,longbibliography]{revtex4-1}
\usepackage{textcomp}
\usepackage{times}
\usepackage{graphicx}
\usepackage{float}
\usepackage{latexsym,amsmath,amssymb,bm,euscript}
\usepackage{color}
\usepackage{subfigure}
\usepackage{epstopdf}
\usepackage[colorlinks=true,linkcolor=blue,citecolor=blue,urlcolor=blue]{hyperref}

\usepackage{soul}
\usepackage[normalem]{ulem}
\usepackage{mathrsfs}
\usepackage{amsmath,amssymb}
\usepackage{lettrine}
\usepackage{xfrac}
\usepackage{xspace}
\usepackage{textcomp}
\usepackage{wasysym}
\usepackage{xr}
\usepackage{lineno}
\graphicspath{{./Figs/}}

\newcommand{\lhy}[1]{{\color[rgb]{.0,.0,.0}{#1}}}

\begin{document}

\title{Continuous transition and gapless roton inside fractional quantum anomalous Hall states}
\author{Hongyu Lu}
\altaffiliation{The two authors contributed equally to this work.}
\affiliation{Department of Physics and HK Institute of Quantum Science \& Technology,
The University of Hong Kong, Hong Kong, China}
\author{Han-Qing Wu}
\altaffiliation{The two authors contributed equally to this work.}
\affiliation{Guangdong Provincial Key Laboratory of Magnetoelectric Physics and Devices, School of Physics, Sun Yat-sen University, Guangzhou 510275, China }
\author{Bin-Bin Chen}
\affiliation{Peng Huanwu Collaborative Center for Research and Education, Beihang University, Beijing 100191, China}
\author{Zi Yang Meng}
\email{zymeng@hku.hk}
\affiliation{Department of Physics and HK Institute of Quantum Science \& Technology,
The University of Hong Kong, Hong Kong, China}
\date{March 2025}

\begin{abstract}
Collective excitations play a vital role in understanding the exotic phases of matter and phase transitions in quantum many-body systems. 
For the first time, we numerically (via exact diagonalization and density matrix renormalization group) report the microscopic realization of a transition from a translationally invariant fractional quantum anomalous Hall (FQAH) state to the same FQAH state with spontaneously broken translation symmetry, by softening the magnetoroton mode (intrinsic collective excitations in such systems) through isotropic interactions in a topological flat-band model. 
At the critical point, the gap of collective neutral excitations closes at finite momentum, while the charge gap remains robust.
This mechanism echoes with the integer quantum Hall crystals and fractional quantum Hall nematics in Landau levels, but exhibits unique features.
Further through criticality analysis, we identify that this non-trivial transition is consistent with the Ising universality class.
Such spontaneous translation symmetry breaking inside the topological ordered FQAH state could serve as a generic scheme in various systems, with experimental implications to the quantum moir\'e materials and the cold-atom systems.
\end{abstract}

\date{\today }
\maketitle

\noindent{\textcolor{blue}{\it Introduction.}---}
Fractional quantum Hall (FQH) effect is one of the most celebrated topics in strongly correlated systems and condensed matter physics~\cite{Tsui1982_FQHE, Laughlin1983_LaughlinState, Haldane1985_Pseudopotential_Exact, Jain1989_composite_fermion, Wen1995_topological_order_edge_excitations, Stormer1999_review_FQH}. Fractional quantum anomalous Hall (FQAH) states or zero-magnetic-field fractional Chern insulators (FCI) are believed as the lattice analog of the FQH states~\cite{Tang2011_flat_chern_band, Sun2011_flat_chern_band, Neupert2011_flat_chern_band, Sheng2011_FQAH_checkerboard_fermion, Regnault2011_FCI}, sharing similarities and having differences~\cite{Regnault2011_FCI, Wu2012_adiabatic_FQH_FCI, Roy2014_band_geometry_fci, Wang2021_geometry_flatband}. 
Recent experimental breakthroughs of discovering FQAH states in two-dimensional moir\'e
materials~\cite{Cai2023_signature_fqah_mote2, Park2023_observation_fqah_mote2,Zeng2023_thermo_evidence_fqah_mote2,Xu2023_Observation_FQAH_tMote2,Lu2024_FQAH_multilayer_graphene} have further motivated studies in understanding both the material-based mechanisms as well as the properties of FQAH states themselves~\cite{Abouelkomsan2024_bandmixing_fqah, Yu2024_FCI_multiband_graphene, Dong2024_QAH_pentalayer_graphene, Reddy2023_FQAH_tTMD}.
Among the important questions, one interesting topic is the interplay of topological orders and charge-density-wave (CDW) orders, i.e., their competition and co-existence~\cite{Reddy2023_FQAH_tTMD, Song2024_phase_transition_moire, Song2024_intertwined_FQAH_CDW, Sharma2024_displacement_field_tMoTe2}.
Recently, the topological states with co-existing CDW orders at zero magnetic field have been generalized as anomalous Hall crystals~\cite{Sheng2024_QAHC, Dong2024_AHC_multilayer_graphene_I, Soejima2024_AHC_multilayer_graphene_II, Zhou2024_FQAH_multilayer_graphene}, with many theoretical discussions~\cite{Dong2024_stability_AHC_multilayer_graphene, Tan2024_ideal_AHC, Patri2024_EQAH_theory} and possible experimental signatures~\cite{Su2025_TEC_graphene, Lu2025_EQAH_graphene}. However, most of these realizations share only integer Hall conductance with filling $\nu\neq\sigma_{xy}$ (possibly suggesting only part of the electrons contribute to the topology while the others form CDW)~\cite{Sheng2024_QAHC, Patri2024_EQAH_theory, Su2025_TEC_graphene, Lu2025_EQAH_graphene}, and the microscopic connection to the symmetric QAH states with the same integer Hall conductance is less clear compared to the original proposal of integer Hall crystals (see below).
Besides, the generic ways to realize FQAH+CDW states at the model level remain elusive. 

\begin{figure}[htp!]
	\centering		
	\includegraphics[width=0.5\textwidth]{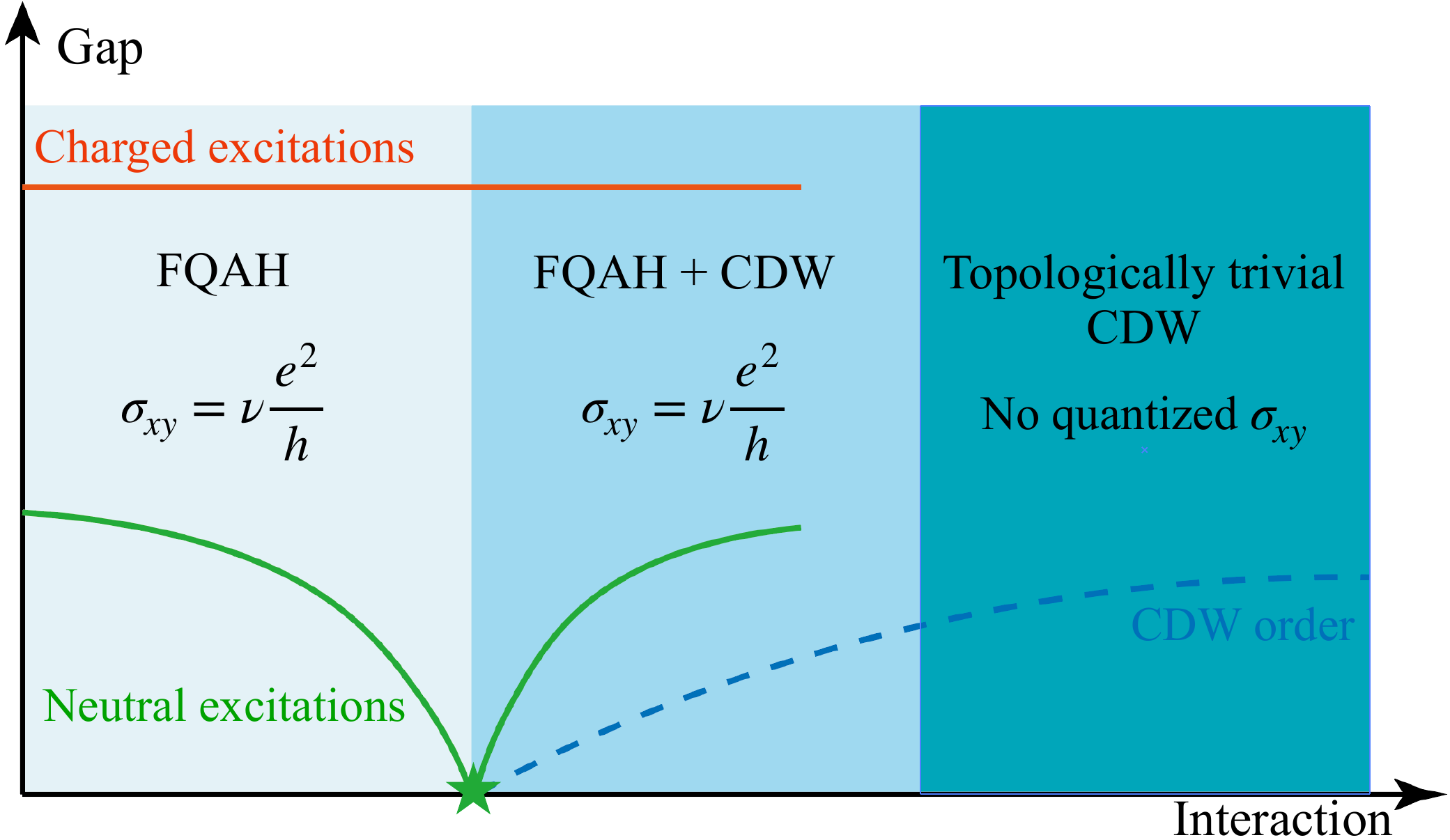}
	\caption{\textbf{Generic phase diagram.} By tuning interaction from a symmetric FQAH state with quantized $\sigma_{xy}=\nu\frac{e^2}{h}$, the system undergoes a two-step transition to a coexisting FQAH+CDW state with the same Hall conductance and moderate CDW order and a topologically trivial CDW state with stronger charge order. The focus is the general mechanism of the first transition. The increasing interaction softens the magnetoroton mode, and at the quantum critical point, the neutral gap continuously closes at finite momentum, which leads to the spontaneously (discrete) translation symmetry breaking, while the charge gap remains open at the quantum critical point as well as the two topological states.}
	\label{fig_fig1}
\end{figure}

The original concept of Hall crystals has been proposed in Landau levels (LLs) for more than 3 decades~\cite{Kivelson1986_ring_exchange_FQH,Halperin1986_compatibility_crystal_QHE, Halperin1989_Hall_crystal}. The idea starts from an integer quantum Hall (IQH) state. When the neutral mode goes soft to gapless while the charge gap remains robust, by tuning the interactions, the ground state becomes a Hall crystal with the same integer Hall conductance and weak CDW order~\cite{Halperin1989_Hall_crystal}. Such a successful demonstration of the integer Hall crystal state is at the mean-field level, and 
although the generalization has been proposed possible for the interaction-dominated FQH states, the realization is still lacking in microscopic models. 
However, there exist many studies showing that the chiral graviton mode (that originates from the intrinsic geometric fluctuations~\cite{Haldane2011_geometric_FQH}) 
in FQH states could go soft in the long-wavelength limit with finite charge gap, leading to the FQH nematics with the same Hall conductivity and broken rotation symmetry (but conserved translation symmetry)~\cite{You2014_FQH_nematics, Regnault2017_FQH_Nematics, Yang2020_nematic_FQH,Pu_FQH_nematics2024}.  

\begin{figure*}[htp!]
	\centering		
	\includegraphics[width=\textwidth]{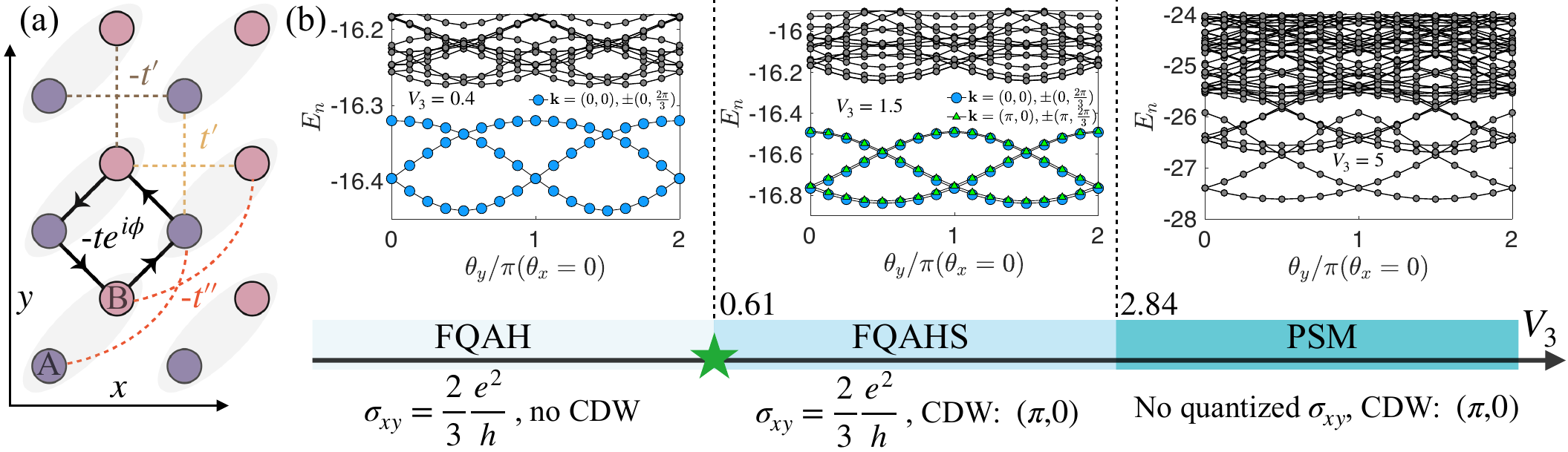}
	\caption{\textbf{Model realization.} 
    (a) The two-band model on checkerboard lattice with $N=N_y\times N_x\times2=3\times2\times2$ sites as an example. A(B) sublattices and different hoppings are denoted by different colors and the arrows represent the directions of the NN loop current. (b) Phase diagram at $\nu=2/3$ filling of the flat Chern band with fixed $V_1=1.1,$ $V_2=1$ and changing $V_3$. The ED spectra from $3\times4\times2$ tori of each phase are shown.
    At small $V_3$, the ground state is a symmetric FQAH state with 3-fold degeneracy and Hall conductivity $\sigma_{xy}=\frac{2}{3}\frac{e^2}{h}$. At intermediate $V_3$, the FQAHS state has the same Hall conductivity as the FQAH state and the co-existing CDW order at $(\pi,0)$. Due to the spontaneously translation symmetry breaking at $V_3=0.61$, the ground-state degeneracy of the FQAHS state is further doubled to 6-fold.
    Further increasing $V_3$, the CDW order becomes stronger and the topological transition leads to the gapless and topologically trivial PSM state.}
	\label{fig_fig2}
\end{figure*}

Roton is also a key elementary excitations in correlated quantum liquids~\cite{Park2024_electronic_roton_dipole}, including the Bose–Einstein condensates (BECs)~\cite{Landau1941_superfluid_helium, Feynman1956_spectrum_liquid_helium}, Fermi liquids~\cite{Godfrin2012_roton_fermi_liquid}, and the FQH states (magnetoroton)~\cite{GMP1986_magneto_roton, Igor2009_dispersion_roton}. FQAH states are believed to be both similar to and beyond the FQH physics in LLs, and recently, their collective excitations including the magnetoroton mode have been intensively discussed in lattice models and moir\'e materials~\cite{Lu_FQAH_thermodynamic2024, Lu2024_vestigial_GBDW_FCI, Long2025_spectra_FCI,Paul2025_collective_modes_moire_FCI, Shen2025_magnetoroton_moire_FCI,Kousa2025_magnetoroton_moire}.
Based on the knowledge of integer Hall crystals~\cite{Halperin1989_Hall_crystal} and FQH nematics~\cite{Pu_FQH_nematics2024}, we here propose a generic scheme for realizing FQAH+CDW state as illustrated in Fig.~\ref{fig_fig1}. Starting from a symmetric FQAH state, if the magnetoroton gap is softened and closed at some finite momentum by tuning interactions while keeping the charge gap finite, the ground state becomes an FQAH+CDW state with the same Hall conductance and spontaneously generated moderate CDW order. 
\lhy{Unlike the integer Hall crystal and nematic FQH states in Landau levels, the neutral excitation of the FQAH+CDW state here is gapped since we consider discrete translational symmetry.}
Stronger CDW order will eventually lead to topologically trivial states~\cite{Halperin1989_Hall_crystal, Pu_FQH_nematics2024}.
As the roton mode is introduced in the superfluid liquid helium and the FQH states are often interpreted as the superfluid of composite bosons~\cite{Zhang1989_field_theory_FQH}, our scenario is analogous to supersolids triggered by the rotons in superfluids~\cite{Tanzi2019_supersolid_roton, Bottcher2019_supersolid_roton, Chomaz2019_supersolid_roton, Zhang2019_supersolid_critical, Alana2023_supersolid_roton}.
Although the intuitive picture is transparent, to the best of our knowledge, no clear realization in a concrete microscopic model for either FQH or FQAH states has been reported, except for the topologically trivial CDWs triggered by rotons~\cite{Kumar2022_continuous_FQH_stripe, Lu2024_vestigial_GBDW_FCI, Shen2025_magnetoroton_moire_FCI}.

This work fills in this knowledge gap. By tuning isotropic interactions at $\nu=2/3$ filling of a topological flat-band model, we find a continuous transition from a symmetric FQAH state to an FQAH+smectic CDW state, dubbed as FQAHS, with the same Hall conductivity. The transition is triggered from the softening of the roton mode while maintaining the charge gap across the transition. Further increasing interaction would lead to a topologically trivial polar smectic metal (PSM) phase sharing non-Fermi-liquid properties~\cite{Emery2000_smectic_metal}.
Through infinite density matrix renormalization group
simulations in the quasi-1D limit, we identify that this continuous FQAH-FQAHS transition is consistent with the 2D Ising universality class as the topology does not change.
As a general example of obtaining FQAH+CDW states from symmetric FQAH states, we believe this work paves the way for better understanding the collective excitations and symmetry-breaking orders in FQAH systems and the possible experimental realizations of FQAH+CDW states.

\noindent{\textcolor{blue}{\it Model, methods and phase diagram.}---}
We consider the topological two-band model on the checkerboard lattice with spinless fermions,
\begin{equation}
	\begin{aligned}
		H =&-\sum_{\langle i,j\rangle}te^{i\phi_{ij}}(c_i^\dagger c^{\ }_j+h.c.)-\sum_{\langle\hskip-.5mm\langle i,j \rangle\hskip-.5mm\rangle}t'_{ij}(c_i^\dagger c^{\ }_j+h.c.)\\
		&-\sum_{\langle\hskip-.5mm\langle\hskip-.5mm\langle i,j \rangle\hskip-.5mm\rangle\hskip-.5mm\rangle} t''(c_i^\dagger c^{\ }_j+h.c.)+\sum_{i,j}V_{ij}n_in_j,
	\end{aligned}
	\label{eq:eq1}
\end{equation}
with : $t=1$ (as the energy unit), $t'_{ij}=\pm 1/(2+\sqrt{2})$, $t''=-1/(2+2\sqrt{2})$ and $\phi_{ij}=\frac{\pi}{4}$, as shown in Fig.~\ref{fig_fig2} (a). The two bands have nontrivial Chern number $C=\pm1$ and the lower Chern band is almost flat with these optimal hopping parameters~\cite{Sheng2011_FQAH_checkerboard_fermion, Sun2011_flat_chern_band}.
We further consider the nearest-neighbor (NN) interaction $V_1$, the next-nearest-neighbor interaction $V_2$, and the third nearest-neighbor interaction $V_3$.
The results in the main text are based on exact diagonalization (ED) and infinite density matrix renormalization group (iDMRG)~\cite{McCulloch2008_iDMRG}. For the iDMRG simulations, we mainly consider the $N_y\times N_x\times2=3\times2\times2$-site unit cells (i.e. 6-leg infinite cylinder) and we keep $D=3000$ $U(1)$ states for most simulations. The maximum truncation error is $\sim10^{-9}$ with good convergence.

\begin{figure*}[htp!]
	\centering		
	\includegraphics[width=\textwidth]{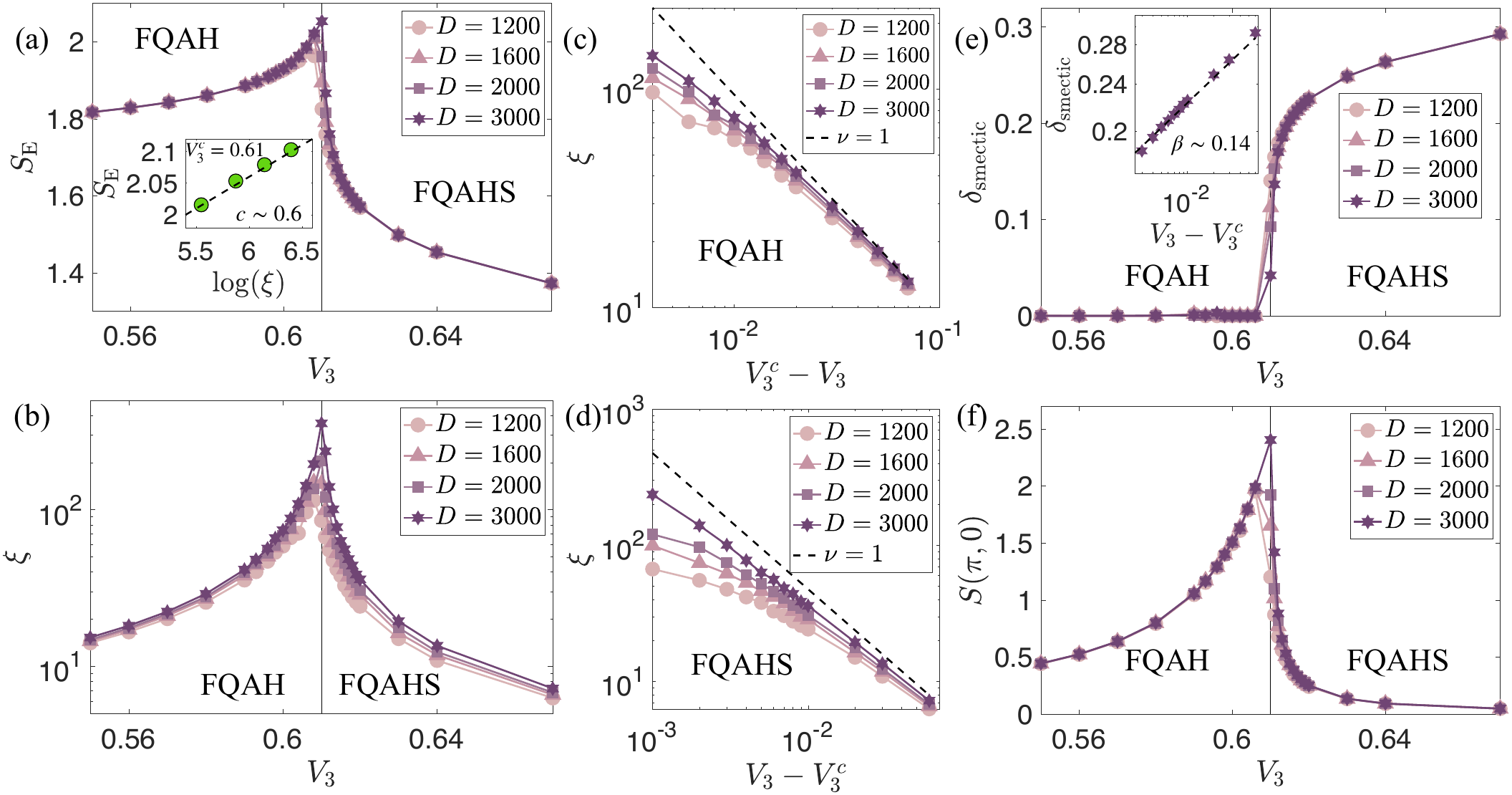}
	\caption{\textbf{Continuous FQAH-FQAHS transition.} iDMRG results at different bond dimensions of a $N_y=3$ cylinder with infinite length  (a) Bipartite entanglement entropy 
    and (b) correlation length as functions of $V_3$, which both diverge at the critical point. 
    \lhy{The inset of panel (a) shows the fitting of central charge at critical point $V_3^c=0.61$ according to $S_\mathrm{E}=\frac{c}{6}\log(\xi)$ and the fitted result from the data points of largest bond dimensions is $c\sim0.6$ (we simulated up to $D=4000$ for this part)}. 
    The power-law divergence of correlation length approaching the critical point at the (c) FQAH and (d) FQAHS sides are shown. When the bond dimension increases, the scalings of correlation length are both approaching $|V_3-V_3^c|^{-\nu}$ with $\nu=1$ of the 2D Ising universality class.
    (e) Smectic CDW order parameter and (f) structure factor at $(\pi,0)$ as functions of $V_3$. The inset of panel (e) in the double-logarithm scale shows the fitting of the order parameter according to $\delta_\mathrm{smectic}\sim (V_3-V_3^c)^\beta$ and the fitted exponent is $\beta\sim0.14$.
    $S(\pi,0)$ that measures the smectic fluctuations also tends to diverge at the critical point and quickly drops after the establishment of smectic order.
    }
	\label{fig_fig3}
\end{figure*}

We focus on the $\nu=2/3$ filling of the flat Chern band. The phase diagram of considering only $V_3$ interaction has been reported~\cite{Lu_FQAHS2024}. The ground state is a Fermi liquid at the non-interacting limit, and is a PSM at the large-$V_3$ limit. The PSM is a non-Fermi liquid with anisotropic transport properties and irrelevant inter-stripe tunnelings~\cite{Emery2000_smectic_metal, Lu_FQAHS2024, Li2024_imaging_luttinger_liquid}. Interestingly, between the Fermi liquid and the PSM states, there is an FQAHS (FQAH + smectic CDW) state.
Although the ground-state and thermodynamic properties of the FQAHS state have been well understood, its origin and the possible connection to any symmetric FQAH state remain unexplored.
In another work at the same $\nu=2/3$ filling, the $V_1-V_2$ phase diagram has been studied~\cite{Lu2024_FQAH_roton_CDW}. There is a symmetric FQAH state with competing interactions, surrounded by different CDW states.
However, whether there exists a direct transition between the symmetric FQAH state and the FQAHS state with coexisting orders in the global phase diagram is still not clear.

In this work, we mainly consider the fixed $V_1=1.1$ and $V_2=1$ and tune $V_3$.  The quantum phase diagram is shown in Fig.~\ref{fig_fig2} (b). Without $V_3$, the ground state is the symmetric FQAH state with $\sigma_{xy}=\frac{2}{3}\frac{e^2}{h}$. Increasing $V_3$, we first find a direct transition to the FQAHS state and then to the PSM state. The energy spectra under twisted boundary conditions from ED simulations are shown for each phase. 
The FQAHS state with the $(\pi,0)$ smectic order has the same Hall conductivity as the FQAH state but the ground-state degeneracy is doubled due to the spontaneously breaking translation symmetry. The PSM state (with stronger smectic order) is gapless and has no quantized $\sigma_{xy}$.
The focus of this work is on the nature of the FQAH-FQAHS transition (which is continuous as shown below), more details and other paths in the global phase diagram are available in the Supplementary Information (SI)~\cite{suppl}. 

\noindent{\textcolor{blue}{\it Continuous FQAH-FQAHS transition.}---}
The bipartite entanglement entropy ($S_\mathrm{E}$) and correlation length ($\xi$) around the first transition are shown in Fig.~\ref{fig_fig3} (a,b), respectively. They both diverge with increasing bond dimensions at the critical point $V_3^c=0.61$, supporting the continuous nature of the transition.
Since the FQAHS state with the same topological order as the FQAH state breaks additional discrete translation symmetry and has a $(\pi,0)$ smectic CDW order and doubled ground-state degeneracy, we expect that it belongs to the Ising universality class. 
In the inset of Fig.\ref{fig_fig3}(a), we extrapolate the central charge at the critical point according to $S_\mathrm{E}=\frac{c}{6}\log(\xi)$ (\lhy{we have considered up to $D=4000$ for this part}). The extrapolated \lhy{$c\sim0.6$} is roughly consistent with the expected $c=0.5$ for the 2D Ising transition (due to the quasi-1D geometry of infinite cylinder in iDMRG simulations). 
We further show the behavior of correlation length as a function of $|V_3-V_3^c|$ when approaching the critical point from both sides in Fig.~\ref{fig_fig3} (c,d) respectively. When the bond dimension of the iDMRG simulations increases, the correlation length scales closer to the power law $\xi\sim |V_3-V_3^c|^{-\nu}$ and the fitted exponent is approaching to $\nu=1$, which further confirms the continuous transition and supports its Ising nature.
As illustrated in Ref.~\cite{Lu_FQAHS2024}, the $(\pi,0)$ smectic order exists in only one sublattice (B sublattice here) in the FQAHS state. Therefore, we define the smectic CDW order parameter as $\delta_\mathrm{smectic}=\frac{2}{N'}\sum_i^{'} (-1)^{x_i} \langle n^\mathrm{B}_{\mathbf{r_i}}\rangle$ 
with the $\sum'$ summation over $N'(=N_xN_y)$ sites. 
Besides, the static structure factor of density-density correlation functions is defined as $S(\mathbf{q})=\sum_je^{-i\mathbf{q}(\mathbf{r_j}-\mathbf{r_0})}(\langle n_j^\mathrm{B}n_0^\mathrm{B}\rangle -\langle n_j^\mathrm{B}\rangle \langle n_0^\mathrm{B}\rangle)$, with $S(\pi,0)$ as the smectic fluctuations.
The smectic CDW order and the smectic fluctuations as functions of $V_3$ are shown in Fig.~\ref{fig_fig3} (e,f), respectively. When approaching the critical point from the FQAH state, the smectic fluctuations gradually increase and peak at $V_3^c$ 
which will eventually diverge with bond dimensions. After the spontaneously established smectic order at the critical point, the smectic fluctuations decrease in the FQAHS state \lhy{due to the explicitly translation symmetry breaking with spatially variant electron distribution}.
In the inset of Fig.~\ref{fig_fig3} (e), we show the smectic order (with bond dimension $D=3000$) as a function of $V_3-V_3^c$ in the double-logarithm scale. We find that the power-law scaling of the order parameter $\delta_\mathrm{smectic}\sim (V_3-V_3^c)^\beta $ with the fitted exponent $\beta\sim0.14$, which is very close to the expected exponent $\beta=\frac{1}{8}$ for the 2D Ising transition.

\begin{figure}[htp!]
	\centering		
	\includegraphics[width=0.46\textwidth]{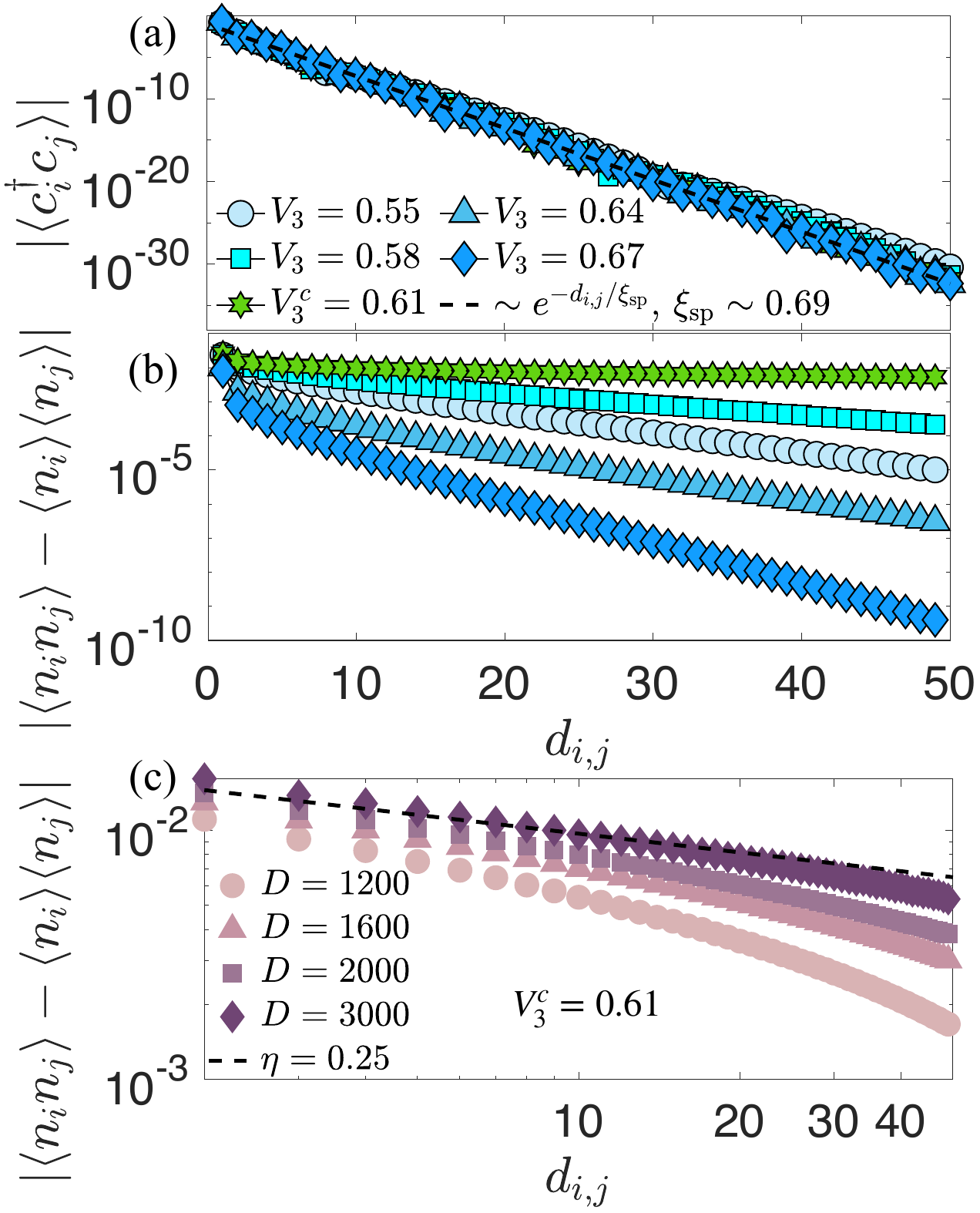}
	\caption{\textbf{Gapless roton with a large charge gap at critical point.}
    The (a) single-particle Green's function and (b) density (neutral) correlation functions at different $V_3$ versus the distance are shown in the semi-log scale and they share the same labels. In panel (a), the exponential decaying behavior of fermion correlations remains across the FQAH-FQAHS transition ($\xi_\mathrm{sp}\sim0.69$), indicating the robust charge gap.
    In panel (b), it is clear that the correlation length of density correlations increases fast when approaching the critical point. 
    (c) The density correlations at the critical point $V_3^c=0.61$ in the log-log scale, where the algebraic decaying is shown. With the increasing bond dimension, the scaling behavior is approaching $\sim d_{i,j}^{-\eta}$ with $\eta=0.25$ for the 2D Ising transition.
    }
	\label{fig_fig4}
\end{figure}

This continuous FQAH-FQAHS transition is a manifestation of the generic scheme illustrated in Fig.~\ref{fig_fig1}. 
We show the single-particle Green's function $|\langle c_i^\dagger c_j^{\ }\rangle|$ and density (neutral) correlation function $|\langle n_in_j\rangle -\langle n_i\rangle \langle n_j\rangle|$ at different $V_3$ in the semi-log scale in Fig.~\ref{fig_fig4} (a,b), respectively. Although the system correlation length (by diagonalizing the transfer matrix~\cite{McCulloch2008_iDMRG}) diverges at the critical point $V_3^c$, we find that the single-particle Green's function decays exponentially across the transition, which means that the charge gap will never close and remains large from FQAH to FQAHS.
The extrapolated correlation length of the exponentially decaying single-particle Green's function is smaller than one lattice spacing ($\xi_\mathrm{sp}\sim0.69$). And we expect the charged excitation remains gapped in the 2D limit (consistent with ED results~\cite{suppl}).
On the other hand, the closer to the critical point, the slower the (neutral) density correlation function decays. The zoom-in behavior of the density correlation function at the critical point is shown with different bond dimensions in Fig.~\ref{fig_fig4} (c). 
When increasing the bond dimension, the decaying behavior of the density correlation function is approaching the power-law scaling versus distance $\sim d_{i,j}^{-\eta}$ with $\eta=0.25$ for the 2D Ising transition.

From these results, we have confirmed that this continuous FQAH-FQAHS transition is indeed due to the condensation of magnetoroton mode that goes soft and closes the neutral gap at $(\pi,0)$. Besides, the charge gap that protects the incompressibility of the FQAH liquid never closes across this continuous transition, such that the fractional Hall conductivity remains across the transition. 
This intriguing scenario is exactly described by the generic scheme (Fig.\ref{fig_fig1}) of obtaining FQAH+CDW states.

\noindent{\textcolor{blue}{\it Experimental implications.}---}
The experimental observations of FQAH states in quantum moir\'e materials have been intensively reported~\cite{Cai2023_signature_fqah_mote2, Park2023_observation_fqah_mote2,Zeng2023_thermo_evidence_fqah_mote2,Xu2023_Observation_FQAH_tMote2,Lu2024_FQAH_multilayer_graphene}. 
However, one overlooked characteristics is that $\sigma_{xy}=\nu \ C_\mathrm{band}$ does not exclude the possibility of the coexisting CDW order or the potential anisotropic transport in FQAH states, as a generic case pointed out here. Therefore in experiments, whether there exists coexisting orders with FQAH states cannot be simply decided by the relation between the quantized Hall conductance and filling, but
needs further experimental detection of the charge order from
other probes such as scanning tunneling spectroscopy.

As the origin of magnetoroton mode is in close analogue to the roton in superfluid~\cite{GMP1986_magneto_roton}, we could further extend the analogy to the supersolids and FQ(A)H+CDW states. 
In the BECs, the supersolids from the softening of the rotons in superfluids have been widely studied.
Such realizations have been reported in ultracold atomic systems
for spin-orbit-coupled BECs~\cite{Li2017_supersolid_spinorbit} and BECs coupled to crossed
optical cavities~\cite{Leonard2017_supersolid_gas_coupled}, where the periodic modulation is induced by periodic optical potentials. 
The rotons could also be softened by tuning the s-wave scattering length in continuous dipolar gas systems~\cite{Tanzi2019_supersolid_roton, Bottcher2019_supersolid_roton, Chomaz2019_supersolid_roton, Zhang2019_supersolid_critical, Alana2023_supersolid_roton}, and by tuning the cavity-mediated long-range interactions in discrete optical lattices~\cite{Mottl2012_roton_cavity,Landig2016_supersolid_longrange_cavity}.

Considering the fast-developing cold-atom, dipolar-gas and quantum-circuit platforms in realising FQ(A)H states~\cite{Yao2013_FCI_dipolar_spin, Grass2018_bosonFQH_laughlin_rydberg, Julian2023photonFQH, Wang2024photonFQH, Shen2025_Laughlin_processor},
we argue that our work provides a generic mechanism and a further strong movitation to explore the anisotropic transport and coexisting CDW orders in such states. 
We notice that the spectroscopic scheme based on two interfering Laguerre-Gaussian beams to detect the magnetoroton and edge excitations in such systems has been proposed~\cite{Binanti2024_spectroscopy_FCI}, and the roton-driven topologically trivial crystals have been reported for such as interacting bosons in the gauge fields~\cite{Mukherjee2022_crystal_bosonic_gas}.
Although the example in our work is by tuning longer-range interactions in a fermionic FQAH state on the lattice, the generic mechanism by softening the roton mode is similar in bosonic and continuous systems, and the similar techniques in the BECs such as the periodic potentials favoring the target CDW momenta could be used.
Besides, we have shown that this roton-driven FQ(A)H-FQ(A)H+CDW transition can be continuous, which is beneficial to the experimental evolutions compared with the first-order transitions (which could lead to the generation of high-energy excitations upon
dynamically crossing the crystallization) as in the superfluid-supersolid evolutions~\cite{Zhang2019_supersolid_critical,Alana2023_supersolid_roton}.
 
\noindent{\textcolor{blue}{\it Discussions.}---}
Our FQAH-FQAHS transition echoes with the origin of the integer Hall crystals and FQH nematics in the Landau-level systems in the sense of the gapless neutral but gapped charge excitations at the critical point and the unchanged Hall conductivity~\cite{Halperin1989_Hall_crystal, Pu_FQH_nematics2024}.
There has been no similar scenarios in the FQAH literature and this work paves the way for better understanding the collective modes in the FQAH systems, which attracts more and more attention recently. 
Besides, this is the first realization of the interaction-driven spontaneously translation symmetry breaking inside the FQ(A)H systems without changing the Hall conductivity (integer Hall crystal is at the mean-field level and the FQH nematics break only rotation symmetry), showing the robustness of topological order against perturbations. We have further provided the numerical evidence of the quantum criticality analysis and contributes to an exotic example of Ising transitions, absent in the literature.

Moreover, our scenario shares another similarity that the interactions finally lead to topologically trivial CDW states with stronger order parameter. While the topological transition from the  integer Hall crystal to Wigner crystal is understood from the further closing single-particle gap~\cite{Halperin1989_Hall_crystal}, the transitions from the FQAHS state here, or the FQH nematics, to topologically trivial and gapless CDW states (PSM here) has not been fully understood~\cite{Pu_FQH_nematics2024}.
In the previous work with only $V_3$ interaction, the ground-state FQAHS-PSM transition is first-order, but the finite-temperature one is possibly continuous~\cite{Lu_FQAHS2024}. Therefore, it is worth investigating whether there exists a continuous FQAHS-PSM transition at zero temperature, which could be a potentially continuous topological transition beyond the traditional Ginzburg-Landau paradigm. We leave this interesting possibility for future work.

A further interesting question is what protects the robust charge gap when the neural gap goes soft. 
In the original proposal of integer Hall crystal, although the tuning interaction is two-body, the four-body interaction is necessary to stabilize the charge gap in that case~\cite{Halperin1989_Hall_crystal}. But in our FQAH-FQAHS transition as well as the FQH-nematic FQH transition~\cite{Yang2020_nematic_FQH, Pu_FQH_nematics2024}, two-body interactions are sufficient. A deeper answer is still elusive. 
Besides, how to soften the graviton mode in FQAH states is also an interesting open question.

In this work, the studied systems do not have exact $C_4$ rotational symmetry due to the choice of geometries. We believe that, in the $C_4$ symmetric systems, the translation symmetry breaking along x and y directions should be degenerate and would further enlarge the ground-state degeneracy in the FQAHS phase. While such system sizes are beyond our current numerical capability, it is interesting for future explorations.
\begin{acknowledgments}
{\it Acknowledgments}\,---\, We thank Kai Sun and Bo Yang for
helpful discussions. We thank Eduardo Fradkin about the Ising nature of this transition and Zhendong Zhang for discussions on ultracold systems. 
We use TeNPy package for the infinite DMRG simulations~\cite{tenpy2018}. HYL and ZYM acknowledge the support from the Research Grants Council (RGC) of Hong Kong (Project Nos. AoE/P-701/20, 17309822, HKU C7037-22GF, 17302223, 17301924), the ANR/RGC Joint Research Scheme sponsored by RGC of Hong Kong and French National Research Agency (Project No. A\_HKU703/22). We thank HPC2021 system under the Information Technology Services and the Blackbody HPC system at the Department of Physics, University of Hong Kong, as well as the Beijing Paratera Tech Corp., Ltd~\cite{paratera} for providing HPC resources that have contributed to the research results reported within this paper. 
H.Q. Wu acknowledge the support from National Natural Science Foundation of China (No. 12474248), GuangDong Basic and Applied Basic Research Foundation (No. 2023B1515120013), Guangdong Fundamental Research Center for Magnetoelectric Physics (Grant No.2024B0303390001), Guangdong Provincial Key Laboratory of Magnetoelectric Physics and Devices (No. 2022B1212010008).
\end{acknowledgments}

\bibliographystyle{apsrev4-2}

\clearpage

\renewcommand{\theequation}{S\arabic{equation}} \renewcommand{\thefigure}{S%
	\arabic{figure}} \setcounter{equation}{0} \setcounter{figure}{0}

\begin{widetext}

	\section{A. ED spectra of fixed $V_1=1.1$ and $V_2=1$}

	\begin{figure}[htp!]
		\centering		
		\includegraphics[width=0.5\textwidth]{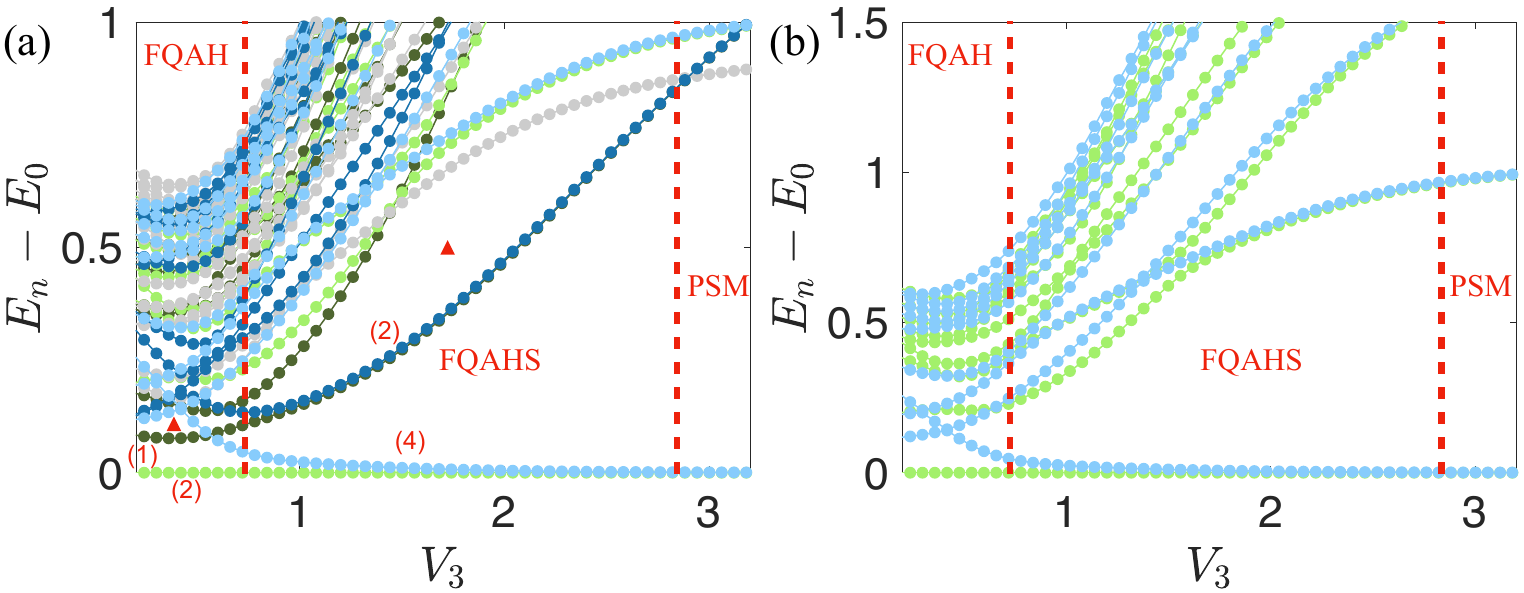}
		\caption{\textbf{Energy spectra with fixed $V_1=1.1$ and $V_2=1$.} These energy levels are obtained from ED simulations of $3\times4\times2$ tori. 
			\lhy{In panel (a), we plot the energy levels from all sectors while we plot only the $(0,\frac{2\pi}{3})$ and $(\pi,\frac{2\pi}{3})$ sectors in panel (b).}
			\lhy{The colors of different momentum sectors are defined in Fig. \ref{fig_figS2}(a).}
			The red dashed lines are the phase boundaries as in the Fig. 2 (b) of the main text. The FQAH state has a 3-fold ground-state degeneracy and the FQAHS state has a 6-fold degeneracy, while the PSM state is gapless. \lhy{We have labelled the number of states near the ground states of the FQAH and FQAHS phases. We also add two small red triangles in panel (a) to represent the neutral gap of the FQAH and FQAHS phases, respectively.} 
			The supporting spectra of each phase under twisted boundary conditions are also shown in Fig. 2 (b) of the main text. The first phase boundary is from the iDMRG results in the main text, while the second phase boundary is determined from the level crossing at $V_3\sim2.84$.
		}
		\label{fig_figS1}
	\end{figure}
	
	\begin{figure}[htp!]
		\centering		
		\includegraphics[width=0.5\textwidth]{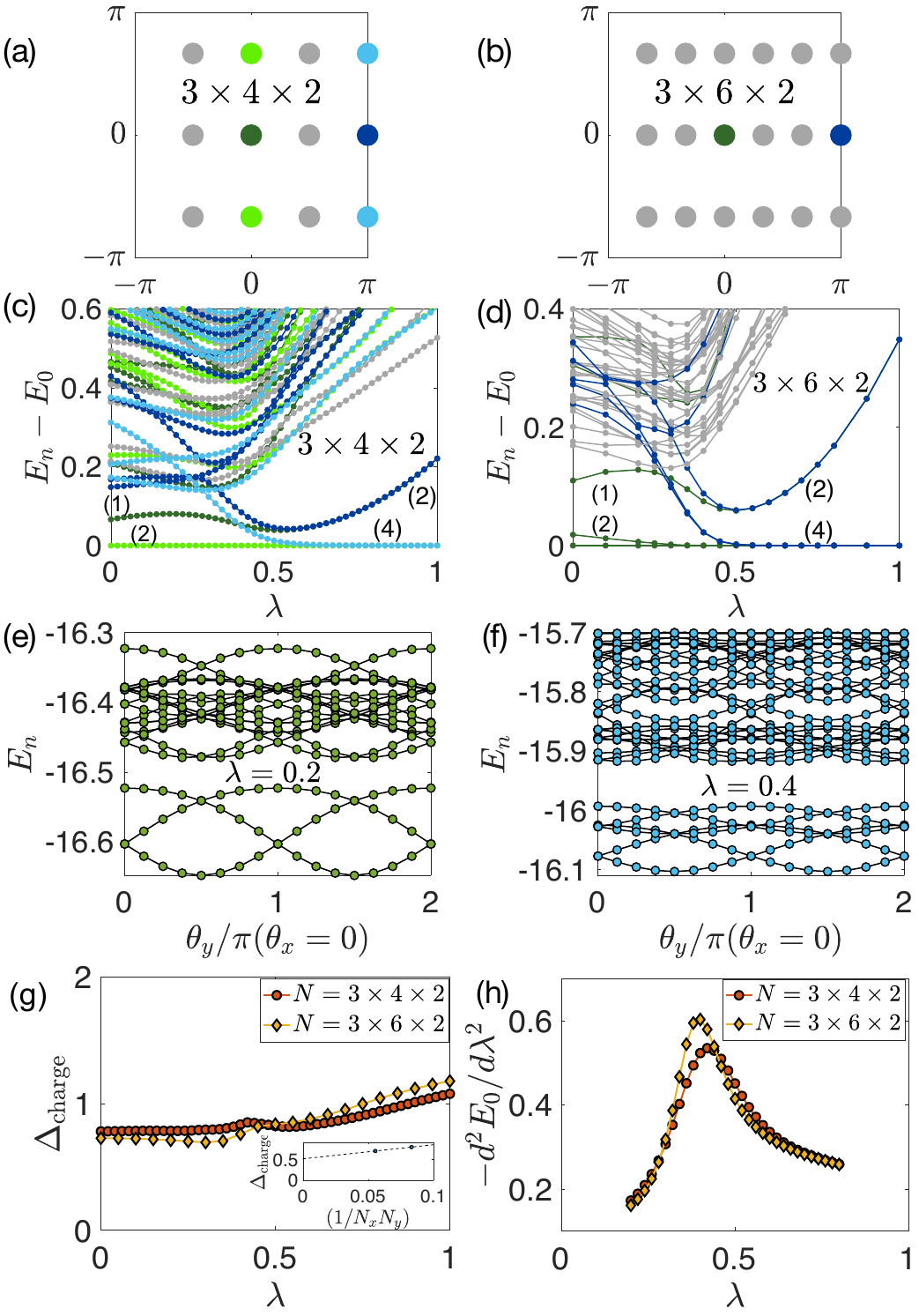}
		\caption{\textbf{ED results of the FQAH-FQAHS transition on a different path.} (a) and (b) show the Brillouin zone of $3\times4\times2$ torus and $3\times6\times2$ torus respectively. The momentum sectors related to the ground states are marked with light/dark green/blue colors, while the others are marked with gray. (c) and (d) show the energy spectra with the changing of $\lambda$ for $3\times4\times2$ torus and $3\times6\times2$ torus respectively, and the colors represent the momentum sectors specified in (a) and (b). In the \lhy{FQAH and FQAHS phases, the number of  degenerate states are labeled}. (e) and (f) show the twisted energy spectrum at $\lambda=0.2$ and $\lambda=0.4$, respectively, exhibiting the 3-fold (6-fold) ground-state degeneracy of FQAH(S) state. The charge gaps with the changing of $\lambda$ for the $3\times4\times2$ \lhy{and $3\times6\times2$ tori} are shown in (g). \lhy{The inset is the extrapolation of charge gap around the transition point, which further supports the nonzero charge gap across the continuous transition.} (h) The second derivative of ground state energy with respective to $\lambda$.
		}
		\label{fig_figS2}
	\end{figure}
	
	In this section, we show the supplementary energy spectra of ED simulations in the same parameter path (fixed $V_1=1.1$ and $V_2=1$) as in the main text.

	Although the spectra of each phase under twisted boundary conditions are shown in Fig. 2 (b), we show the evolution of spectra (without twisting the boundary conditions) when tuning $V_3$ in Fig.\ref{fig_figS1} (a).
	The first phase boundary is from the iDMRG results in the main text, and it is also consistent with the ED spectra. Although it is under periodic boundary conditions, it is still clear that the neutral gap closes and re-opens around $V_3\sim0.61$, and we will explain more about the evolution of ground states in the next section. The second phase boundary is determined from the level crossing at $V_3\sim2.84$.
	Note that, in our ED simulations, the repulsive interactions are written as $V_{ij}\sum_{ij}(n_i-1/2)(n_j-1/2)$, which leads to some simultaneous shift of the exact values but not the relative values of the energies.
	\lhy{We also show the spectra from only the $(0,\frac{2\pi}{3})$ and $(\pi,\frac{2\pi}{3})$ sectors in Fig. \ref{fig_figS1}(b). Although they should be exactly degenerate in the thermodynamic limit for both ground states and excited states, there will be energy splitting due to finite-size effect, which is more obvious near the FQAH-FQAHS transition point and also more obvious for excited states.}
	
	\section{B. The FQAH-FQAHS transition on a different path }
	In this section, we show another parameter path to demonstrate the FQAH-FQAHS transition from ED simulations.
	To study the transition in the global three dimensional parameter space, we define a parameter $\lambda$ such that $V_1$, $V_2$ and $V_3$ interactions are functions of $\lambda$. 
	More specifically, we set $V_1=1.5(1-\lambda)$, $V_2=1$, and $V_3=1.5\lambda$.
	In Fig.~\ref{fig_figS2} (a,b), the momentum sectors of the utilized $3\times4\times2$ and $3\times6\times2$ tori in the Brillouin zone of the two-site unit cells are shown, where those containing the ground state sectors are marked with dark/light green/blue color, while the others are in gray. The energy spectrum from $\lambda=0$ to $\lambda=1$ under periodic boundary conditions of the two tori are shown in Fig.~\ref{fig_figS2} (c,d), respectively, and the definitions of the colors used here are the same as in Fig.~\ref{fig_figS2} (a,b). 
	In the $3\times4\times2$ case, when $\lambda$ is small, the 3-fold ground states of the FQAH state are from $(0,0)$ and $(0,\pm\frac{2\pi}{3})$ sectors, while those in the $3\times6\times2$ case are all from the  $(0,0)$ sector.
	When approaching the critical point around $\lambda\sim0.34$ and in the process of forming the smectic order, there are another three states from $(\pi,0)$ and $(\pi,\pm\frac{2\pi}{3})$ sectors in the $3\times4\times2$ case and all from the $(\pi,0)$ sectors in the $3\times6\times2$ case, merging into the original ground states. 
	Therefore, there exists 6-fold degeneracy in the FQAHS ground states, and we show the energy spectra of a $3\times4\times2$ torus under twisted boundary conditions at $\lambda=0.2,\ 0.4$, in Fig.~\ref{fig_figS2} (e,f) respectively. \lhy{While the states with $k_x=0$ and $\pi$ are exactly degenerate in the thermodynamic limit, at $\lambda=0.4$ (close to the critical point $\lambda\sim0.34$), the energies of states with $k_x=\pi$ are relatively higher than those with $k_x=0$ due to the finite-size effect.}

	The charge gaps defined as $\Delta_\mathrm{charge}=E(N, N_\mathrm{e}+1)+E(N, N_\mathrm{e}-1)-2E(N, N_\mathrm{e})$
	in the  $3\times4\times2$ \lhy{and $3\times6\times2$} tori along the same path of changing $\lambda$ are shown in Fig.~\ref{fig_figS2} (g) \lhy{and the inset shows the extrapolation of charge gap around the transition point}, which is consistent with the iDMRG results in the main text that although the neutral excitation gap is closed at the critical point, the charge excitation remains gapped. 
	The second derivatives of the ground state energy (taken from the state with lowest energy) with respective to the tuning parameter $\lambda$ for the two system sizes are shown in Fig.~\ref{fig_figS2} (h), which also supports that this FQAH-FQAHS transition is continuous, which can also be seen from the smooth merging of the ground states around the transition point in Fig.~\ref{fig_figS2} (c) and (d).

	\section{C. Supplementary information of the FQAHS state}
	\begin{figure}[htp!]
		\centering		
		\includegraphics[width=0.3\textwidth]{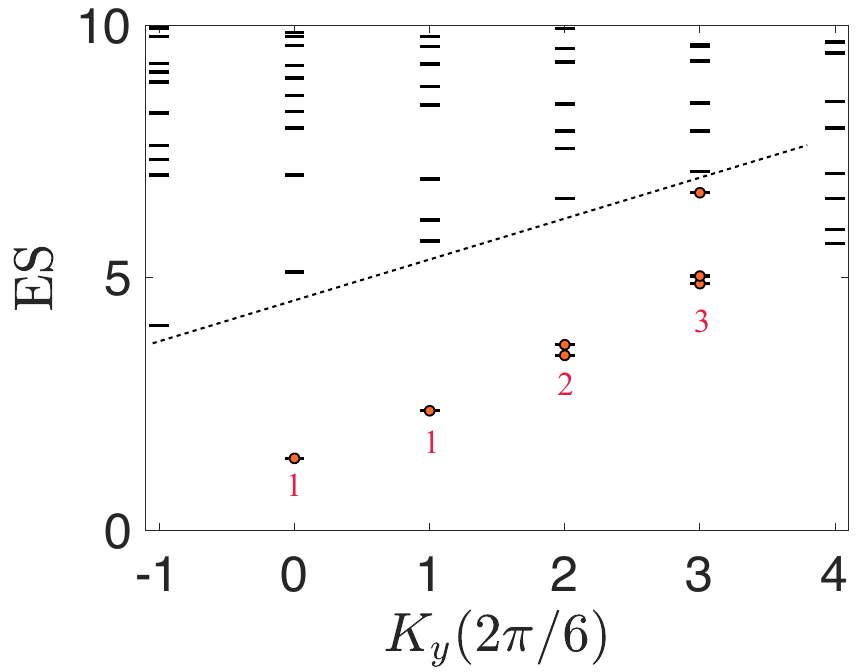}
		\caption{\textbf{Momentum-resolved entanglement spectrum (ES) of the FQAHS phase.} \lhy{This is obtained from iDMRG simulations with $N_y=6$ at $\nu=2/3$, and $V_1=1.1,\ V_2=1,\ V_3=0.8$. The characteristic counting $\{1, 1, 2, 3, ...\}$ of the edge
				conformal field theory is observed, confirming the topological nature of this state.}
		}
		\label{fig_figS3}
	\end{figure}
	\lhy{To provide additional topological properties of the FQAHS phase, we compute its momentum-resolved entanglement spectrum in iDMRG simulations. Although the simulation becomes very hard around the critical point when increasing system size, we can obtain converged result at $N_y=6$ deep inside the gapped FQAHS state (we consider $V_1=1.1,\ V_2=1,\ V_3=0.8$). As shown in the new Fig. \ref{fig_figS3},  the CFT edge theory counting $\{1,1,2,3,...\}$ is observed~\cite{Li2008_ES}, confirming the topological nature of the state.}
	
	\section{D. The FQAH-FQAHS-PSM transitions at $\nu=1/3$}
	\begin{figure}[htp!]
		\centering		
		\includegraphics[width=0.5\textwidth]{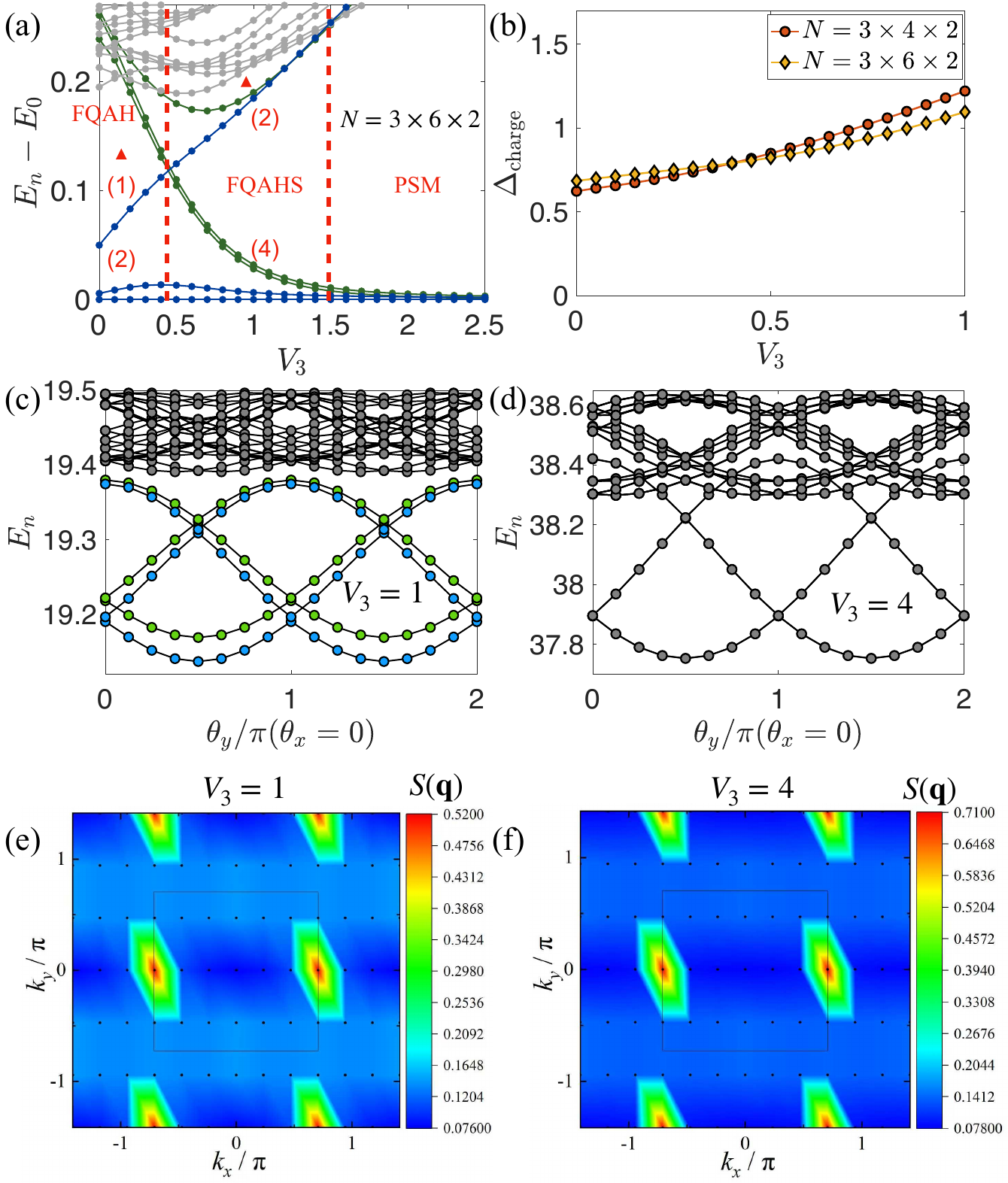}
		\caption{\textbf{ED results at $\nu=1/3$ with fixed $V_1=4$.} \lhy{(a) The ED spectra from the $N=3\times6\times2$ torus. The number of ground states and the neutral gap (red triangle) of the FQAH and FQAHS are labelled. The red dashed lines represent the phase boundaries.
				(b) The nonzero charge gap across the FQAH-FQAHS transition.
				(c-d) The spectral flow of the FQAHS ($V_3=1$) and PSM ($V_3=4$) states, respectively.
				(e-f) The structure factors of the FQAHS ($V_3=1$) and PSM ($V_3=4$) states, respectively.}
		}
		\label{fig_figS4}
	\end{figure}
	
	\lhy{In this section, we check if similar FQAHS state could occur at $\nu=1/3$ filling of the lower flat band.
		As shown in the new Fig. \ref{fig_figS4}, we consider $\nu=1/3$ of the lower flat band and start from $V_1=4$ where the ground state is the symmetric FQAH state. The spectra from the $3\times6\times2$ torus is shown in  Fig. \ref{fig_figS4}(a), when gradually turning on the $V_3$ interaction, the original neutral gap above the 3-fold ground states from (0,0) sector (of the FQAH state) closes. After the gap re-opening, there are totally 6-fold quasi-degenerate states with 3 more from the $(\pi,0)$ sector.
		The gapped spectral flow of the FQAHS state is shown in  Fig. \ref{fig_figS4} (c). As we only consider 1 parameter path here, the gap could be larger in other regions of the parameter space. 
		When further increasing $V_3$, we observe similar PSM state with gapless spectra as shown in  Fig. \ref{fig_figS4}(d). The structure factors of the FQAHS and PSM are shown in  Fig. \ref{fig_figS4}(e,f), respectively, with sharp peaks at $(\pi,0)$.
		We also show the charge gap across the FQAH-FQAHS transition from two system sizes in  Fig. \ref{fig_figS4}(b), suggesting that the charge gap would not close across this transition as well.}
	
\end{widetext}

\end{document}